\documentclass[aps,pra,twocolumn,superscriptaddress]{revtex4}

\usepackage{amsfonts,epsfig}
\usepackage{latexsym}
\usepackage{amsmath}
\usepackage{amssymb}
\usepackage{amsfonts}
\usepackage{amsthm}
\usepackage{mathrsfs}
\usepackage{natbib}
\usepackage{color,verbatim,graphics}
\usepackage{psfrag}

\newcommand{\ba}{\begin{eqnarray}}
\newcommand{\ea}{\end{eqnarray}}
\newcommand{\ban}{\begin{eqnarray*}}
\newcommand{\ean}{\end{eqnarray*}}
\newcommand{\ket}[1]{|#1\rangle}
\newcommand{\bra}[1]{\langle#1|}

\newcommand{\ceil}[1]{\left\lceil#1\right\rceil}

\begin{document}

\title{Limiting the complexity of quantum states: a toy theory}

\author{Valerio Scarani}
\affiliation{Centre for Quantum Technologies, National University of Singapore, 3 Science drive 2, Singapore 117543}
\affiliation{Department of Physics, National University of Singapore, 2 Science Drive 3, Singapore 117542}
%etc.

%\date{\today}

%______________________________________________________________________ ABSTRACT

\begin{abstract}

This paper discusses a restriction of quantum theory, in which very complex states would be excluded. The toy theory is phrased in the language of the circuit model for quantum computing, its key ingredient being a limitation on the number of interactions that \textit{each} qubit may undergo. As long as one stays in the circuit model, the toy theory is consistent and may even match what we shall be ever able to do in a controlled laboratory experiment. The direct extension of the restriction beyond the circuit model conflicts with observed facts: the possibility of restricting the complexity of quantum state, while saving phenomena, remains an open question.

\end{abstract}

\maketitle

%______________________________________________________________________ Article

\section{Introduction}

Since the early days, physicists have faced the task of explaining why quantum theories don't seem to matter for the objects of our everyday experience. Nowadays, there are two well-established routes for an effective quantum-to-classical transition: decoherence and coarse graining. The possibility of more radical routes has been rather dormant for some years, confined to the small community studying spontaneous collapse models, but is now getting new momentum, since it is very often mentioned as a motivation for the experimental tests of quantum effects with large objects \cite{Bassirev13,optorev14}. This paper shares in the idea that something radical may happen in the quantum-to-classical transition, but explores a different path than spontaneous collapse.

The \textit{size} of the system is the first possible critical parameter that comes to mind. Size definitely sets strong limitations to the activity of the physicist: none of us will ever write down the most general state of a few hundred of qubits, since $2^{265}\approx 10^{80}$ is the estimated number of atoms in the observable universe. Beyond these practical considerations, however, sheer size does not say much about degrees of freedom. For instance, the observation of an interference fringe for matter waves herald superposition of a \textit{single} degree of freedom, the centre of mass, irrespective of how massive the system actually is. Though not logically absurd,  in face of the evidence collected so far it is hard to conceive that a single degree of freedom be not described by quantum physics. For such delocalisation experiments, some invoke decoherence due to gravity: this would wash out spatial interference indeed, but would not erase the effects of superposition of other degrees of freedom --- a dead cat looks pretty much the same as a sleeping one from a gravitational perspective, so a superposition of the two possibilities would not be decohered by that mechanism \footnote{I use the well known cat metaphor, being aware that it is even doubtful whether ``to be alive/dead" can be defined as states, see e.g. B.-G. Englert, Eur. Phys. J. D \textbf{67}, 238 (2013).}.

If quantum theory has to fail somewhere on the route from microscopic entities to the scale of our everyday experience, rather than size, it may be preferable to invoke some definition of \textit{complexity}. In recent years, the possibility that very complex states predicted by quantum theory may not exist in nature has been put forward to cast doubts on the feasibility of quantum computing. In an attempt to bring the debate onto more rational ground than simple suspicion, Aaronson proposed a quantitative way of defining which families of multipartite states deserve the label of ``complex" \cite{Aaronson04,CLS15}. However, to my knowledge nobody has tried to describe how a theory of interacting systems would look like, that matches quantum theory in ``simple" situations and deviate from it in ``complex" ones \footnote{Several proposals have been put forward on how one may have both quantum and irreducibly classical objects. These works are infinitely more advanced than what I am going to propose here and still encounter problems [for a recent example, see: M.J.W. Hall, M. Reginatto, C.M. Savage, Phys. Rev. A \textbf{86}, 054101 (2012)]. Besides, to my knowledge, they do not have a parameter that could be varied to move from one side to the other.}. This paper describes such a toy theory.

%For objects made of many constituents, quantum theory allows for superpositions that must involve many degrees of freedom, irrespective of which decomposition is chosen \footnote{The fact that some states are intrinsically multimode is well known in quantum optics. An example can be found already for two photons. Indeed, consider a two photon state that, in some mode decomposition, reads $\frac{1}{2}(a^{\dagger 2}+b^{\dagger 2})\ket{0}$. This state can be re-written without entanglement as $c^\dagger_+c^\dagger_-\ket{0}$ with $c_{\pm}=\frac{1}{\sqrt{2}}(a^{\dagger}\pm b^{\dagger})$. But it still involves two modes, and one cannot find a single mode $d$ such that the same state would be written $\frac{1}{\sqrt{2}}d^{\dagger 2}\ket{0}$.}. 

\section{The toy theory}
\label{sec:toy}

\subsection{Definition and general properties}

The toy theory is defined in the framework of the circuit model for quantum computation. It is assumed that each degree of freedom is a qubit, and that the possible gates are any single-qubit rotations and CNOTs for the interactions. As widely known, this is one of the possible sets that leads to universal quantum computing. The toy theory achieves its goal by \textit{limiting the number of two-qubit interactions that each qubit can undergo}. No constraints are imposed on the other two elements of the circuit model, single-qubit rotations and single-qubit destructive measurements.

Specifically, let us say that each qubit can take part in at most $K$ CNOT gates, either as control or as target. If a CNOT gate involves a qubit that has already undergone $K$ previous ones, it simply won't act and the state will be left unchanged.

Before attempting a more quantitative discussion of what this toy theory can do, and in particular to which extent it reproduces quantum theory and when it starts departing from it, let me make a few qualitative observations.

The first important observation is that \textit{the toy theory is a restriction on ordinary quantum theory and can be simulated within it}. For instance, one may append to each qubit a \textit{register} with $K+1$ orthogonal states, initialised in the state $k=0$. Then, the action of two-qubit gates is conditioned on the value $k$ of the register: if $k<K$ for both qubits, the gate acts as a CNOT, and the register of both qubits is shifted as $k\rightarrow k+1$; if one of the qubits has $k=K$, the gate acts as the identity and the registers are unchanged. This observation shows that \textit{the toy theory cannot have any of the inconsistencies that often plague attempts at modifying quantum theory}. In particular, the no-signalling feature is respected, as should be obvious from the fact that the action of each gate depends only on the state of the two interacting systems.

Besides, as soon as $K\geq 2$, the theory allows the creation of the GHZ state of arbitrarily many qubits. This state is generally considered as the paradigm of ``macroscopic superposition" \cite{FD12}. Thus, \textit{no restriction is induced on the number of qubits that can become entangled}, only on the type of many-qubit entanglement that can be created.

The most striking departure of the toy theory from accepted postulates is the fact that \textit{the dynamics of isolated systems is no longer fully reversible}. In fact, the dynamics can be reversed if and only if each of the qubits has participated in at most $K/2$ CNOTs (assuming that circuits are optimally designed). Notice that, even when it becomes impossible to revert to a previous state, a pure state remains a pure state. This is very different from the kind of irreversibility that would be called ``collapse". In fact, just like in the circuit model of quantum computing, measurement is going to be treated as a primitive: this toy theory does not have a measurement theory and is making no attempt at solving a hypothetical measurement problem, which exists only in some classes of interpretations --- more generally, \textit{the toy theory is agnostic as to whether states should be interpreted in an epistemic or an ontic way} \cite{Leifer}.

\subsection{Why me? [M. Balotelli, public communication]}

This is the toy theory we are going to work with. Of course, once the Pandora box is open, one may think of many variations, and indeed \textit{it is the main goal of this paper to stimulate such a reflection}. Here are some thoughts on the matter:

\begin{itemize}

\item The option that the interaction may become trivial only if \textit{both} qubits have $k=K$ is of some interest and will be quickly mentioned later (Section \ref{sec:beyond}). This variation would share the same qualitative features just described, and for a given value of $K$ would recover more of quantum theory.

\item Instead of restricting the interactions, one may restrict directly the amount of entanglement or of complexity, defined according to one's favourite measure, and see what it implies on the dynamics. Aaronson's idea of \textit{tree size complexity}, mentioned above, would fit here. I'll discuss in the technical part (paragraph \ref{kts}) why the toy theory discussed here does not have a strong link with Aaronson's tree size complexity.

\item In the same vein, for some time I toyed with the idea of restricting simply \textit{the number of qubits that can be entangled}, but I was stopped by what I consider a major obstacle. Consider the case, where an entangled state of the largest allowed number of qubits has been prepared: one must be able to describe what happens if one of those qubits (call it Q) is brought in interaction with a new one. Now, whichever \textit{ad hoc} recipe one may invent, it must rely on information that is \textit{not locally available}: indeed, Q must know to be part of a how-many-qubit entangled state, which is obviously not the same as knowing with how many qubits it has interacted directly.
\end{itemize}

\section{Tuning the value of $K$}
\label{sec:tunek}

One of the nice features of the toy theory is the tuneable parameter $K$. For $K=1$, the theory allows qubits to get entangled only by pairs, after which one can only rotate and measure them: besides violating Bell inequalities, there is little one can do with that. For $K=2$, the theory already allows entangling arbitrarily many qubits, at least in the GHZ state. For $K\rightarrow\infty$ the toy theory is the standard circuit model for quantum computing.

This raises the curiosity of knowing what the theory can and cannot do for a given value of $K$. This section is devoted to a first exploration of this question.

\subsection{The threat for public key encryption}
\label{ss:shor}

As a simple warm up, let us see what value of $K$ one should set for Shor's algorithm to be able to break current public key encryption schemes. Clearly, the \textit{circuit depth} provides an upper bound on $K$, which would be tight if at least one of the qubits participates in a CNOT at every step.

Now, in order to factor an integer of $n$ bits, Shor's algorithm requires at most $2n+3$ logical qubits, with a circuit depth whose leading term is $32 n^3$ \cite{beauregard,FH04}. In order to compromise the RSA encryption based on $n=2048$, one needs $K\gtrsim 2^{38}$ (assuming that each logical qubit is encoded in one physical qubit; a larger bound would be obtained if one considers fault-tolerant architectures).

\subsection{Reproducing all of QM up to $n$ qubits}

A natural set of interesting values of $K$ are the thresholds $K_n$, defined as the minimal value of $K$ for which the toy theory can create \textit{all} pure states of $n$ qubits, starting from the product $\ket{0}^{\otimes n}$. Bounds on $K_n$ (which, we recall, is a number of CNOTs \textit{per qubit}) can be inferred from bounds on the \textit{total} number of CNOT gates required to create an arbitrary $n$-qubit state, which have been the object of several studies.

A \textit{lower bound} on $K_n$ is readily obtained. Indeed, by counting parameters, one finds that at least $K_n^{total}\geq \frac{1}{2}(2^n-n-1)$ CNOT gates in the whole circuit are needed if one wants to create any $n$-qubit pure state (see e.g. Section II of \cite{PB11}). Thus each qubit must take part in at least $K_n^{total}/(n/2)$ such gates, whence
\ba
K_n&\geq& K_n^{L}=\ceil{\frac{2^n-1}{n}}-1\,.
\ea
Tight circuits are known for $K_2=K_2^{L}=1$ and $K_3=K_3^{L}=2$ \cite{ZGG08}. Already for $n=4$, the lower bound is $K_4^{L}=3$ but the best known circuit has 4 CNOTs on all qubits (Fig.~1 of \cite{PB11}, improved by an observation made in Appendix E of \cite{ICKHC15}).

As we said above, an \textit{upper bound} on $K$ is given by the circuit depth: in this case, we need the depth of a circuit capable of producing all states of $n$ qubits from product ones. To date, the best known such bound is the one provided by Plesch and Brukner \cite{PB11}:
\ba
K_{n}&\leq& K_{n}^{U}\,=\,\ceil{c\,2^n}
\label{knu1}\ea with $c=\frac{23}{48}$.

The gap between these lower and upper bounds grows with $n$, since $K_{n}^{U}/K_{n}^{L}\approx \frac{n}{2}$. Just to have some numerical values to fix the ideas, we note that $102\leq K_{10}\leq 491$, $52428\leq K_{20}\leq 502443$. Also, even if it is obvious that such must be the case, it is impressive to note the difference between implementing the Shor's algorithm and realising the most general state of the same number of qubits: referring to the numbers cited in paragraph \ref{ss:shor}, $K_{2\times 2048+3}\sim 2^{4100}\gg 2^{38}$. 

\subsection{$K$ and tree-size complexity}
\label{kts}

Just a short paragraph to stress that there is no direct relation between Aaronson's tree-size complexity and the states that are forbidden in the toy theory once a value of $K$ is chosen. The reason is twofold. First, usually the question one asks in the context of tree size is not whether a state is ``highly complex" for a given number of qubits: rather, one asks if the tree size increases super-polynomially with $n$ for a given family of states. Second, tree size complexity does not imply high circuit complexity: Aaronson proved super-polynomial tree size for the so-called sub-group states and (up to a technical conjecture) for the states that appear in Shor's algorithm; both families can be realised with a polynomial number of CNOT gates.

Now, the toy theory constrains circuit complexity, the constraints on state complexity being only a consequence. Since it would make no sense to allow $K$ to depend on the number $n$ of qubits that are considered part of the ``system", the number of CNOT gates in the circuit cannot increase more than linearly with $n$ and is bounded by $Kn/2$.

\section{The quantum-to-classical transition}
\label{sec:qtoc}

As we said above, the toy theory does not even try to solve the measurement problem, assuming there is one. However, it would be pleasant to check that classicality can be recovered along both usual routes. The route through \textit{coarse graining} is certainly viable, because it amounts at introducing some imperfections in the measurement devices or in the devices that store the measurement results. The route through \textit{decoherence} is more tricky, since the toy theory restricts the interactions, and these restrictions should not depend on whether the qubit is interacting with another qubit in the ``system" or with a qubit of the ``environment". In the extreme case, once a qubit has interacted through $K$ CNOTs, its state is frozen up to local rotations. So, the theory does put limits to decoherence. However, we are going to see on an example that decoherence is very effective and ``consumes" a small number of interactions.

To stay within the circuit model, we consider a bath of qubits, i.e. a large number of uncorrelated qubits each in the state $\sigma_T=p\ket{0}\bra{0}+(1-p)\ket{1}\bra{1}$. The creation of the bath itself is trivial within the toy theory: with just one CNOT acting on each qubit, one can prepare two qubits in the entangled state $\sqrt{p}\ket{00}+\sqrt{1-p}\ket{11}$; one is taken to form part of the bath, the other one is not used.

The effectiveness of decoherence can be seen by using a \textit{thermalisation channel} tailored for the circuit approach \cite{scaranithermo}. Each qubit of the system interacts sequentially with one qubit of the bath after another. The interaction is a partial swap parametrised by $\cos\phi$, such that $\cos\phi=1$ means no interaction while $\cos\phi=0$ represents the case in which the system qubit is dumped as such into the bath and replaced by a bath qubit in a single step. Intermediate values of $\cos\phi$ lead to a more or less gentle thermalisation process; thus a large value of $\cos\phi$ may be preferable on physical grounds.

The key point is that the convergence of state of the system $\rho^{(m)}$ towards $\sigma_T$ is exponential in the number of interactions $m$. In terms of trace distance, the channel gives $
||\rho^{(m)}-\sigma_T||_1 \leq (\cos\phi)^m$ \footnote{The exact expression is $||\rho^{(m)}-\sigma_T||_1=c^m\,\sqrt{|k_0\lambda^n|^2+(d_0-p)^2c^{2n}}$ where $c=\cos\phi$, $d_0$ and $k_0$ are the parameter of the initial state, and $\lambda$ involves the phase factors of the interaction and is such that $|\lambda|\leq 1$.}. Thus, the state of each system qubit can be thermalised to a precision $||\rho^{(m)}-\sigma_T||_1\leq\varepsilon$ after interacting with at most $m=\log\varepsilon/\log(\cos\phi)$ qubits of the bath. Since each partial swap can be implemented with 3 CNOTs, thermalisation can be guaranteed if one allows
\ba
K_T(\varepsilon, \phi) &\leq& 3\,\frac{\log\varepsilon}{\log(\cos\phi)}
\ea CNOT gates to act on the system qubit. The channel under study was not chosen as to minimise $K$, so this result is only an upper bound.

For the sake of the example, with $\cos\phi=0.99$ one finds $K_T(\varepsilon)\approx 687 \log_{10}\frac{1}{\varepsilon}$: that is, each digit of precision in thermalisation costs some 700 additional CNOTs, which is of the order of $K_{n=11}$. In other words, if we choose $K$ so that quantum theory is exact for at least a couple of tens of qubits, thermalisation comes basically for free.

\section{Beyond the qubit circuit model}
\label{sec:beyond}

This sections contains a few remarks about the (im)possibility of extending the current toy theory beyond the circuit model with qubits.

\subsection{Circuits with higher-dimensional systems}

Staying first with circuit models, one may ask how to generalise the prescription of the toy theory if the elementary systems are taken to be higher-dimensional. One may be initially tempted to fall back by exploiting the obvious mathematical fact that any system of finite dimension $d$ can be mapped into $\log_2 d$ qubits. However, if we take that path, \textit{even single-qudit unitaries would no longer come for free}. The reason is clear: once the state of the qudit is encoded in qubits, a generic single-qudit rotation involves the preparation of entangled state; thence, it require CNOTs.

It seems therefore more appropriate to consider that each degree of freedom has a proper dimension $d$ (it is a local prescription). Single-system unitaries and measurements can then always be considered free resources, while the number of interactions can be bounded by $K_d$. The choice of $K_d$ is not indifferent. If $K_d$ grows enough fast with $d$, for instance, the toy theory may lead to an intriguing ``reverse simulation" of quantum computing: all the algorithms, which would be off-boundary for qubits because of the value of $K_{d=2}$, would become feasible by mapping the qubits into qudits.

\subsection{Measurement-based quantum computation}

With $K=3$ CNOT gates (actually, phase gates) per qubit, one can create the cluster state on a hexagonal lattice, which is known to be universal for quantum computing in the measurement-based model \cite{vdn06}. The price to pay is in number of qubits. Consider an algorithm that, in the circuit model, acts on $n$ qubits and requires $m$ CNOT gates in total: in order to implement that algorithm one needs a cluster state of $O(nm)$ qubits, whose preparation requires just as many phase gates: the total number of required CNOT gates basically does not change between the two models.

Thus, if one stays with the prescription of the current toy theory but introduces algorithms in the measurement-based model, and if the number of qubits is in unlimited supply (as tacitly assumed above), the current toy theory becomes trivial for $K\geq 3$. Surely, in order to obtain all states of $n$ qubits, one would need a cluster state of $O(n\,2^n)$ qubits, and as noticed above this number becomes comparable to the number of atoms in the whole visible universe when $n$ is a couple of hundreds.

\subsection{Interactions}

For an extension to any physical system, the finite set of discrete-time interactions of the circuit model will have to be replaced by a continuous set of continuous time interactions. In this view, a reasonable generalisation may pass through imposing a limit on the \textit{action} that can be spent in interactions, i.e. a bound of the type $Jt\leq K\hbar$ with $J$ the strength of the interaction Hamiltonian and $t$ the time allowed. I am doubtful whether it makes sense to embark in such a  generalisation: if one just looks at atoms, there is no indication that electrons stop interacting with the nucleus after some time. 

The previous observation may just be the last word. For the sake of the reasoning, though, I find it interesting to bring up another issue that may pass more easily unnoticed. The circuit model, and in fact a lot of simple quantum exercises, leave aside the fact that \textit{single-qubit rotations and measurements are also interactions}: the first with an ``external field", the second with a ``measurement device". If the spirit of the theory is to limit interactions, one has to deal with this issue. It would certainly undesirable to have a theory in which, after some time, some systems cannot be measured anymore because they have exhausted their capacity of interaction!

Many solutions can be envisaged, but if we stay with the toy theory as defined in the text, it seems to me that they all pass through endowing ``measurement devices" with special features --- something that may be pleasant or unpleasant according to one's preferred philosophy, but ultimately one would have to identify these special features in real devices. For instance, one may define ``measurement devices" by the power of resetting the register $k$ to the initial value 0. This would ensure that every system can always be measured, and after being measured it can undergo a whole new series of interactions (if the measurement was not destructive, that is).

It is at this point that one may see the benefit of the other version of the toy theory mentioned at the end of Section \ref{sec:toy}, in which interaction fails to happen only of \textit{both} interacting systems have exhausted their capacity of interaction. Indeed, both an ``external field" and a ``measurement device" are almost by definition ``macroscopic" objects (and so would be a thermal bath, to come back to the case of decoherence discussed above). In this other version of the toy theory, then, single-qubit rotations and measurements, and decoherence, would remain virtually unaffected, as long as one requires macroscopic objects to have a very large $K$ as compared to that of qubits. May a variation of this point of view save the observation about electron-nuclei interactions made above?

\section{Conclusion}

I have presented a toy theory that reproduces quantum physics in several conditions but prevents the creation of states that are too complex. This is done by limiting the number $K$ of interactions that a system may undergo. For any value of the parameter $K$, the toy theory is in principle falsifiable. The possibility of simulating the toy theory with standard quantum mechanics guarantees that the theory is not plagued by flagrant inconsistencies. The only radical departure from accepted principles is the prediction of irreversible dynamics for isolated systems (though in a way that pure states remain pure, thus remaining compatible with epistemic interpretations).

The foundations of this idea would become much more solid if one could find a plausible reason to limit the number of interactions: however, I don't see any such reason; worse, as noted above, it looks rather contradicted by observed facts. Other more subtle modifications may lead to theories that differ from quantum theory while agreeing on all observed facts --- or maybe a limitation of complexity will just prove inconsistent, just as all attempts of taming quantum correlations in a classical way have been doomed to failure so far \footnote{For a review of the failed attempts after Bell's pioneering work, see Chapter 5 of V. Scarani, Acta Physica Slovaca \textbf{62}, 347 (2012).}. The current toy theory should not be asked to do much beyond its scope, which was to provide a first example of what a theory with those features would look like.

\section*{Acknowledgments}

This work was done mostly in my spare time, but in general my research is funded by the Singapore Ministry of Education and by the National Research Foundation of Singapore. I am grateful to Cai Yu, Roger Colbeck and Le Huy Nguyen for suggesting relevant references. My thoughts on the difference between size and degrees of freedom where triggered by a remark of Berge Englert some years ago; I shared with him a previous version of this draft, guessing that he would find it pointless and tell me why; he did not let me down. More encouraging feedback came from Gerald Milburn, Michael Hall and Bill Munro, who stressed in different ways that this text raises interesting concerns.

\end{document}